\documentclass[twocolumn, prd, letterpaper, nofootinbib,longbibliography]{revtex4}

\usepackage{graphicx}
\usepackage{mathrsfs}
\usepackage{amsmath}
\usepackage[colorlinks=true,citecolor=blue,urlcolor=cyan,linkcolor=black]{hyperref}

\newcommand{\be}{\begin{eqnarray}}

\newcommand{\ee}{\end{eqnarray}}

\begin{document}

\title{Cross-correlation of the Polarizations of the 21-cm and Cosmic Microwave Backgrounds}

\author{Lingyuan Ji}
\email{lingyuan.ji@jhu.edu}
\affiliation{Department of Physics and Astronomy, Johns Hopkins University\\ 3400 North Charles Street, Baltimore, MD 21218, United States}
\author{Selim C. Hotinli}
\email{shotinl1@jhu.edu}
\affiliation{Department of Physics and Astronomy, Johns Hopkins University\\ 3400 North Charles Street, Baltimore, MD 21218, United States}
\author{Marc Kamionkowski}
\email{kamion@jhu.edu}
\affiliation{Department of Physics and Astronomy, Johns Hopkins University\\ 3400 North Charles Street, Baltimore, MD 21218, United States}

\begin{abstract}
	The polarization of the 21-cm radiation from the epoch of reionization arises from Thomson scattering of 21-cm photons from free electrons and provides information that complements that from the intensity fluctuation. Previous work showed that a direct detection of this signal will be difficult, and hinted that the signal might be enhanced via correlation with other tracers. Here, we discuss the cross-correlation between the cosmic microwave background (CMB) polarization and the 21-cm polarization. We treat reionization using an analytical model with parameters calibrated by semi-numerical simulations. We then derive the cross-correlation angular power spectrum using the total-angular-momentum formalism. We also provide a  noise analysis to test against two closely related, but subtly different, null hypotheses. First, we assume no reionization as a null hypothesis, and determine how well this null hypothesis could be ruled out by an observed 21cm-CMB polarization correlation. Second, we determine how well the null hypothesis of no 21-cm polarization can be ruled out by seeking the cross-correlation, assuming reionization is established from the CMB.  We find that the first question could be answered by a synergy of ambitious next-generation 21-cm and CMB missions, whereas the second question will still remain out of reach. 
\end{abstract}

\maketitle

\section{Introduction}

The redshifted 21-cm line of the hydrogen hyperfine transition provides both a spatially and temporally resolved image of the baryons' growth in inhomogeneity, collapse, heating, and reionization \cite{Pritchard:2011xb, Furlanetto:2006jb, Morales:2009gs}. A wide range of cosmological and astrophysical information can be derived from it. Previous work focused primarily on the intensity signal, showing its potential power in constraining fundamental physics \cite{Weltman:2018zrl, CosmicVisions21cm:2018rfq}, star and galaxy formation \cite{Mellema:2012ht}, and  intergalactic medium \cite{Fan:2006dp}.

More can be learned from the polarization signal. The dominant contribution to the 21-cm polarization arises in the same way as the polarization of the cosmic microwave background (CMB). In the reionized Universe, the Thomson scattering between of a radiation quadrupole from a free electron generates a linear polarization. This effect has first been explored in Refs.~\cite{Babich:2005sb} and \cite{Li:2021wkb}. Ref.~\cite{Babich:2005sb} estimates the strength of this signal by assuming a relatively simple model of reionization, and claims that it can be detected by Square Kilometre Array (SKA) with a one-month observation time. Ref.~\cite{Li:2021wkb}, with refined reionization modeling via {\tt 21cmFAST}, concludes that this signal is smaller than the SKA sensitivity. However, Ref.~\cite{Li:2021wkb} also points out that the signal may still be detected by cross-correlation with other probes.

In this paper, we discuss the cross-correlation between the 21-cm polarization and the CMB polarization. The CMB polarization, produced also by the Thomson scattering, is generated in roughly two different epochs---one right before recombination, the other after reionization. The latter, as we will show, gains contribution from density perturbations within a certain comoving-wavelength range that also give rise to the 21-cm polarization. This generates a cross-correlation. By correlating the well-established CMB signal with a to-be-detected signal, we enhance the sensitivity to the target signal.

To assess the prospects to detect this cross-correlation, we evaluate the ability of future measurements to answer two related but subtly different questions.  In the first, we determine the possibility to rule out the null hypothesis of no reionization from this cross-correlation alone.  In the second, we assume that reionization has been well established from the CMB-polarization measurement and ask whether the cross-correlation can be detected under this assumption.  We conclude that the amplitude of the cross-correlation is large enough to distinguish it from the null hypothesis of no reionization, but it cannot be detected under the null hypothesis of a well established detection of the reionization in the CMB polarization.

The rest of the paper is organized as follows. In Section~\ref{sec:theory}, we present the theoretical calculation of the cross-correlation signal. In Section~\ref{sec:forecast}, we analyze the detection prospect of this signal by next-generation CMB and 21-cm observations, while pointing out potential hurdles. We end in Section~\ref{sec:conclusion} with some concluding remarks.

\section{Theory}
\label{sec:theory}

\begin{figure}
    \centering
    \includegraphics[width = \linewidth]{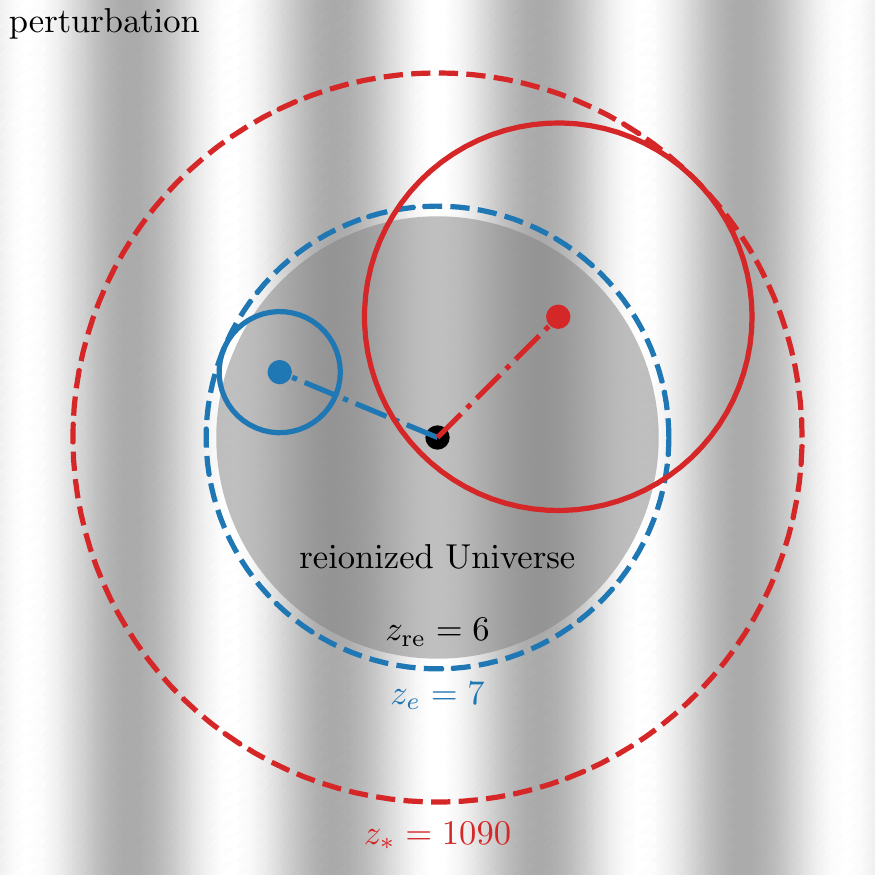}
    \caption{A comoving-space sketch of the physics described in Section~\ref{sec:theory}. The observer is located at the central black point. The blue dashed ring represents the emission shell of the 21-cm photons for the \emph{observer}. The reionized Universe (gray disk) contains free electrons. One electron (blue point) sees an intensity anisotropy, resulting from the perturbations (gray plane-wave pattern) at the \emph{electron's} emission shell of the 21-cm photons (solid blue ring). This then produces a linear polarization seen by the observer (along the blue dot-dashed line). Note that the electron's emission shell is different from, but tangent to, that of the observer's. The similar mechanism for the CMB polarization is depicted using the corresponding red elements, while the dashed and the solid rings should be interpreted as the last-scattering surfaces seen by the observer and the electron, respectively. This sketch is drawn \emph{to scale} according to the labeled redshifts.}
    \label{fig:comoving-plot}
\end{figure}

The physical picture is illustrated schematically in Fig.~\ref{fig:comoving-plot}, which shows the observable Universe in comoving coordinates. As described in the figure caption, we start by discussing the remote intensity anisotropy in Section~\ref{sec:intensity-quadrupole}; we then move to compute the induced linear polarization in Section~\ref{sec:polarization}. These steps are done for both the CMB polarization and the 21-cm polarization. Next, we use the results from these previous steps to calculate the cross-correlation angular power spectrum in Section~\ref{sec:angular-power-spectrum}. Finally, we proceed to describe the ``bubble model'' for the 21-cm fluctuations in Section~\ref{sec:reionization-model}, and its correlation with matter fluctuations. We also discuss how the bubble-model parameters are calibrated to semi-numerical simulations using {\tt 21cmFAST}.

\subsection{Intensity quadrupole}
\label{sec:intensity-quadrupole}

We first discuss the 21cm-intensity quadrupole. Here, for simplicity, we take the post-heating limit, where the spin temperature $T_S$ is very large compared to the temperature $T_\gamma$ of the CMB photons, and neglect the effect of redshift-space distortion. Under these assumptions, the differential brightness temperature $T_{21}$ of the redshifted 21-cm line is then proportional to the neutral-hydrogen density $n_{\rm HI}$ at the hydrogen-photon interaction. Since the Thomson scattering is achromatic, it suffices to only account for the fractional perturbation $\delta_{21}\equiv T_{21}/\bar T_{21}-1$, where $\bar T_{21}$ is the spatial average of $T_{21}$.

 Since the Thomson scattering is achromatic, the linear polarization of a given frequency $\nu_o$ we observe is \emph{only} due to the scattering of those 21-cm photons emitted at redshift $z_e = \nu_e/\nu_o -1$, and is \emph{independent} of the redshift of the scattering. Here, $\nu_e=1420\,\text{MHz}$ is the proper frequency of the hydrogen line. Thus the relevant intensity-anisotropy pattern (i.e.\ responsible for the observed polarization at frequency $\nu_o$) seen by an electron at $(\vec x, \eta)$ from direction $\hat u$ is
\begin{equation}
    \Theta_{21, \nu_o}(\hat u; \vec x, \eta) \equiv \delta_{21}[\vec x + (\eta-\eta_e) \hat u, \eta_e],
\end{equation}
where $\eta_e$ is the conformal time at redshift $z_e$. We define the quadrupole tensor
\begin{equation}
t^{21,\nu_o}_{ab}(\vec x, \eta)\equiv \int d^2\hat u\, (3 u_a u_b - \delta_{ab})\Theta_{21,\nu_o}(\hat u; \vec x, \eta).
\end{equation}
It will be useful later to express $t^{21,\nu_o}_{ab}$ using $\delta_{21}$ in the Fourier space. With detailed derivation given in Appendix~\ref{app:quadrupole-fourier}, we write the result here,
\begin{equation}\label{eqn:t_r21-fourier}
    \tilde t^{21,\nu_o}_{ab}(\vec k, \eta) = -12\pi j_2[k(\eta-\eta_e)]\left(\hat k_a \hat k_b -\frac{\delta_{ab}}{3}\right)\tilde \delta_{21}(\vec k, \eta_e).
\end{equation}
Here, $j_n(x)$ is the spherical Bessel function; $\delta_{ab}$ is the Kronecker delta symbol; $k\equiv |\vec k|$ and $\hat k\equiv \vec k/k$ are the magnitude and direction of the wavevector $\vec k$.

Next, we discuss the the CMB intensity quadrupole. The polarization of CMB is generated both right before recombination and after reionization. However, only the latter contributes significantly to the cross-correlation with the 21-cm polarization. To see this, we note that, for a given free electron, the intensity quadrupole---thus the resulting polarization---is dominated by perturbations with a wavelength roughly the size of the electron's last-scattering-surface radius. For those free electrons right before recombination, this size is roughly the duration $\Delta \chi_* \sim 15\,{\rm Mpc}$ of recombination (assuming recombination at $z_*=1090$ with a duration of $\Delta z_* = 80$). So, the recombination polarization pattern will peak at multipole $l \sim \chi_*/\Delta\chi_* \sim 10^3$, where $\chi_* \sim 1.4\times 10^4\, {\rm Mpc}$ is the comoving distance to the last-scattering surface. At these high multipoles, the Limber approximation \cite{Lemos:2017arq} implies that the 21-cm polarization, coming from a comoving distance different from $\chi_*$, will have very limited correlation with the recombination CMB polarization.

The relevant CMB intensity quadrupole, seen by the free electrons in the reionized Universe, can then be simply calculated using the Sachs-Wolfe effect. Similar to the 21-cm case, the intensity-anisotropy pattern,
\begin{equation}
    \Theta_{\rm CMB}(\hat u; \vec x, \eta) \equiv -\frac13 \Phi[\vec x + (\eta-\eta_*) \hat u, \eta_*],
\end{equation}
determines the quadrupole tensor
\begin{equation}
t^{\rm CMB}_{ab}(\vec x, \eta)\equiv \int d^2\hat u\, (3 u_a u_b - \delta_{ab})\Theta_{\rm CMB}(\hat u; \vec x, \eta).
\end{equation}
Here, $\Phi$ is the conformal-Newtonian-gauge gravitational potential, and $\eta_*$ is the conformal time at recombination. This relation can be written in Fourier space, following the derivation in Appendix~\ref{app:quadrupole-fourier}, as
\begin{equation}
    \tilde t^{\rm CMB}_{ab}(\vec k, \eta) = 4\pi j_2[k(\eta-\eta_*)]\left(\hat k_a \hat k_b - \frac{\delta_{ab}}{3}\right) \tilde \Phi(\vec k, \eta_*).
\end{equation}
We can re-express this equation in terms of the matter-density perturbation $\delta$ at conformal time $\eta_e$ by realizing that
\begin{equation}
    \tilde\Phi(\vec k,\eta)=\frac{3 H_0^2 \Omega_m}{2k^2a(\eta)} \tilde \delta(\vec k, \eta)
\end{equation}
is constant deep in the matter-dominated era, giving \cite{Ji:2020uro}
\begin{equation}\label{eqn:t_CMB-fourier}
    \tilde t^{\rm CMB}_{ab}(\vec k, \eta) = \frac{6\pi H_0^2 \Omega_m}{k^2 a(\eta_e)} j_2[k(\eta-\eta_*)]\left(\hat k_a \hat k_b - \frac{\delta_{ab}}{3}\right) \tilde \delta(\vec k, \eta_e).
\end{equation}
Here, $H_0$ is the current Hubble parameter; $\Omega_m$ is the matter-density parameter; $a(\eta)$ is the scale factor at conformal time $\eta$. We formally switch the time slice from $\eta_*$ to $\eta_e$, so that we only need to refer to the equal-time correlation between $\delta$ and $\delta_{21}$ at the conformal time $\eta_e$ of the 21-cm emission.

\subsection{Polarization}
\label{sec:polarization}

The linear polarization tensor, as a function of position $\hat n$ on the sky, is
\begin{equation}
    P_{AB}(\hat n) \equiv \frac{1}{\sqrt2}
    \begin{pmatrix}
    Q(\hat n) & U(\hat n)\sin \theta\\
    U(\hat n)\sin \theta & -Q(\hat n) \sin^2 \theta
    \end{pmatrix},
\end{equation}
where, $Q(\hat n)$ and $U(\hat n)$ are the Stokes parameters measured in the local coordinate frame $\{\hat \theta, \hat \phi\}$.  The components of $P_{AB}$ are given in the coordinate chart $\{\theta,\phi\}$, indicated by the capital subscripts $AB$. The $\sin \theta$-related factors follow from the fact that the chart $\{\theta,\phi\}$ is orthogonal but not orthonormal. Later, we will mainly use the \emph{Cartesian} components $P_{ab}$ of this polarization tensor, which can be obtained by embedding the unit sphere into the three dimensional space and performing a general coordinate transformation between $\{\theta,\phi\}$ and $\{x,y,z\}$. In both representations, the polarization tensor is symmetric ($P_{AB}=P_{BA}$ and $P_{ab}=P_{ba}$) and trace free ($g^{AB}P_{AB}=0$ and $g^{ab} P_{ab}=0$, with $g_{AB}={\rm diag}\{1,\sin^2\theta\}$ and $g_{ab}=\delta_{ab}$). The embedding also implies that $P_{ab}$ is transverse ($\hat n_a P_{ab}=0$).

Both the CMB and the 21-cm polarization are produced by Thomson scattering between free electrons and unpolarized incident radiation possessing a quadrupolar intensity anisotropy. This indicates that, for the probe $X={\rm CMB}\ \text{or}\ 21(\nu_o)$, we have
\begin{equation}\label{eqn:P-t}
	P^X_{ab}(\hat n) = \frac{\sqrt2}{16\pi}\int g(\eta)d\eta\, \Pi_{ab,ij}(\hat n) t^X_{ij}[\hat n (\eta_0-\eta), \eta],
\end{equation}
where the integral starts from $\eta_*$ for $X={\rm CMB}$ and $\eta_e$ for $X=21(\nu_o)$, to the current conformal time $\eta_0$, and will be omitted from now on. Here, $g(\eta) \equiv e^{-\tau} d\tau/d\eta$ is the photon-visibility function, where the Thomson-scattering optical depth is defined as
\begin{equation}
	\tau(\eta) \equiv \int_{\eta}^{\eta_0} \frac{\sigma_T n_b x_e(\eta') }{a(\eta')^2}\, d\eta'. 
\end{equation}
Here, $\sigma_T$ is the total cross section of Thomson scattering; $n_b$ the comoving number density of baryons; $x_e(\eta)$ the mean ionization fraction [see Eq.~(\ref{eq:mean_reio})]. Note that $\sigma_T n_b = 2.307 \times 10^{-5}(\Omega_b h^2)\, {\rm Mpc}^{-1}$ in terms of the baryon-density parameter $\Omega_b$. The tensor projector is
\begin{equation}
	\Pi_{ab,ij}(\hat n) \equiv \mathcal P_{ai}(\hat n) \mathcal P_{bj}(\hat n) - \frac12 \mathcal P_{ij}(\hat n) \mathcal P_{ab}(\hat n),
\end{equation}
where $\mathcal P_{ij}(\hat n) \equiv \delta_{ij}-\hat n_i \hat n_j$. Eq.~(\ref{eqn:P-t}) must take this form because the integrand is the only possible way to project a three dimensional quadrupole tensor $t_{ij}$ to a polarization tensor $P_{ab}$ that is simultaneously symmetric, trace free, and transverse. The pre-factor in Eq.~(\ref{eqn:P-t}) can be determined by taking $\hat n=-\hat z$ and comparing to Eqs.~(2.10) in Ref.~\cite{Kosowsky:1994cy}.

\subsection{Angular power spectrum}
\label{sec:angular-power-spectrum}

The statistical property of $P^X_{ab}(\hat n)$ is encoded in the angular power spectrum $C^{XY}_J \equiv \langle (P^X_{JM})^* P^Y_{JM}\rangle$, where the E-mode spherical-harmonic expansion coefficients are defined as
\begin{equation}\label{eqn:P_JM}
	P^X_{JM} \equiv \int d^2 \hat n\, \left[Y^{\text{TE}}_{(JM)ab}(\hat n) \right]^* P^X_{ab}(\hat n).
\end{equation}
using the tensor spherical harmonics $Y^{\text{TE}}_{(JM)ab}(\hat n)$ defined in Ref.~\cite{Dai:2012bc}. There exists another family of B-mode expansion coefficients, obtained by projection onto $Y^{\text{TB}}_{(JM)ab}(\hat n)$ (defined also in Ref.~\cite{Dai:2012bc}), that will be zero under our assumption of no primordial gravitational waves.

The next step is to plug Eq.~(\ref{eqn:P-t}) into Eq.~(\ref{eqn:P_JM}) and calculate the angular power spectrum. This task is immensely simplified by the use of total-angular-momentum (TAM) waves defined in Ref.~\cite{Dai:2012bc}, whose notation we shall now follow. Assuming only primordial scalar perturbation, the symmetric and trace-free tensor $t^X_{ab}$ can be expanded as
\begin{equation}\label{eqn:t-expansion}
	t^X_{ab}(\vec x, \eta) = \sum_{kJM} t^X_{kJM}(\eta) \left[4\pi i^J \Psi^{k,\rm L}_{(JM)ab}(\vec x)\right]
\end{equation}
in terms of the L-mode tensor TAM waves. Here, $\sum_k$ is a shorthand for $\int k^2 dk/(2\pi)^3$. The projection in Eq.~(\ref{eqn:P-t}) is then simplified by the property
\begin{equation}\label{eqn:TAM-projection}
	\Pi_{ab,ij}(\hat x)\Psi^{k,\text{L}}_{(JM)ij}(\vec x) = R^\text{L,TE}_J(kx) Y^\text{TE}_{(JM)ab}(\hat x)
\end{equation}
of the TAM waves. Here the radial function $R^\text{L,TE}_J(kx)$ can be inferred from Eq.~(38) of Ref.~\cite{Inomata:2018vbu}.  It will also be helpful to expand the matter-density perturbation $\delta$ and the 21cm-temperature perturbation $\delta_{21}$ in terms of the scalar TAM waves. To simplify the writing of parallel equations, we formally rename
\begin{equation}
    \delta_{\rm CMB}(\vec x, \eta) = \delta(\vec x, \eta)\quad \text{and}\quad \delta_{21,\nu_o}(\vec x, \eta) = \delta_{21}(\vec x , \eta),
\end{equation}
and exclusively refer to the symbols on the left-hands sides when we abstractly express the probe using $X$ and $Y$. Now, the expansions can be compactly written as
\begin{align}\label{eqn:delta-expansion}
	\delta_X (\vec x, \eta) &= \sum_{kJM} \delta^X_{kJM}(\eta) \left[ 4\pi i^J \Psi^k_{(JM)}(\vec x) \right].
\end{align}
Similar to the Fourier-space relations, Eqs.~(\ref{eqn:t_r21-fourier}) and (\ref{eqn:t_CMB-fourier}), $t^X_{kJM}$ can also be related to $\delta^X_{kJM}$. With details given in Appendix~\ref{app:tam-transfer}, we list the result here,
\begin{equation}\label{eqn:t-delta}
    t^X_{kJM}(\eta) = \mathcal T_X(k; \eta, \eta_e) \delta^X_{kJM}(\eta_e).
\end{equation}
Here the quadrupole transfer functions $\mathcal T_X$ are defined as
\begin{align}
    \mathcal T_{21,\nu_o}(k; \eta, \eta_e) &\equiv 4\pi \sqrt6 j_2[k(\eta-\eta_e)];\\ 
    \mathcal T_{\rm CMB}(k; \eta, \eta_e) &\equiv -\frac{2\pi\sqrt6 \Omega_m H_0^2}{k^2 a(\eta_e)} j_2[k(\eta-\eta_*)].
\end{align}

With this setup, we can plug Eq.~(\ref{eqn:t-delta}) into Eq.~(\ref{eqn:t-expansion}), and then the result into Eq.~(\ref{eqn:P-t}) with the aid of Eq.~(\ref{eqn:TAM-projection}). Finally, the orthonormality of $Y^{\text{TE}}_{(JM)ab}(\hat n)$ implies that
\begin{equation}
	P^X_{JM} = \frac{\sqrt2}{4}i^J \int \frac{k^2 dk}{(2\pi)^3}\, \mathcal I^X_J(k; \eta_e) \delta^X_{kJM}(\eta_e) ,
\end{equation}
where the polarization transfer functions $\mathcal I^X_J$ (not to be confused with $\mathcal T_X$) are defined as
\begin{equation}
    \mathcal I^X_J(k; \eta_e) \equiv \int g(\eta)d\eta\, R^\text{L,TE}_J[k(\eta_0-\eta)] \mathcal T_X(k; \eta, \eta_e).
\end{equation}
The angular power spectrum then evaluates to
\begin{equation}\label{eqn:angular-power-spectrum}
	C^{XY}_J = \frac{1}{8} \int \frac{k^2 dk}{(2\pi)^3}\, P_{XY}(k; \eta_e) \mathcal I^X_J(k;\eta_e)  \mathcal I^Y_J(k;\eta_e),
\end{equation}
where the power spectrum is defined such that the products of the TAM coefficients have the expectation values
\begin{equation}
    \langle [\delta^X_{kJM}(\eta_e)]^* \delta^Y_{k'J'M'}(\eta_e) \rangle = \delta_{kk'} \delta_{JJ'} \delta_{MM'} P_{XY}(k;\eta_e).
\end{equation}
Here $\delta_{kk'}$ is a shorthand for $(2\pi)^3\delta(k-k')/k^2$. Note that this definition coincides with the canonical definition using Fourier coefficients, as detailed in Ref.~\cite{Dai:2012bc},
\begin{equation}
    \langle \tilde \delta^*_X(\vec k, \eta_e) \tilde \delta_Y(\vec k', \eta_e) \rangle = (2\pi)^3 \delta(\vec k-\vec k')P_{XY}(k;\eta_e).
\end{equation}

\subsection{Reionization model}
\label{sec:reionization-model}

The question now reduces to obtaining $P_{XY}$, the cross- and auto-spectrum of the matter-density perturbation $\delta$ and the 21cm-temperature perturbation $\delta_{21}$. We employ a simple but effective ``bubble model'' provided in Ref.~\cite{Wang:2005my}, with its parameters calibrated to semi-numerical simulations via {\tt 21cmFAST}. We emphasize that here we are, in a sense, only using the bubble model as a fitting template of the simulation results, so we avoid interpreting the fitted values of the model parameters in a physical way. This approach is similar to that used in Ref.~\cite{Li:2021wkb} (i.e.\ assuming a specific bias scaling outside the $k$-range of the simulation).

Ref.~\cite{Wang:2005my} models the ionization-fraction field via a collection of (possibly overlapping) fully ionized bubbles. The bubbles are uniformly distributed across space, and have a log-normal radius $R$ distribution $P(R)$,
\begin{equation}
    P(R) = \frac{1}{R \sigma_{\ln R}\sqrt{2\pi}} \exp\left[-\frac{(\ln R - \mu_{\ln R})^2}{2\sigma_{\ln R}^2}\right],
\end{equation}
where $\mu_{\ln R}$ and $\sigma_{\ln R}$ are the mean and the standard deviation of $\ln R$, respectively. Note that the mean bubble radius is then $\langle R \rangle = \exp(\mu_{\ln R} + \sigma_{\ln R}^2/2)$.

\begin{figure*}
    \centering
    \includegraphics[width = \linewidth]{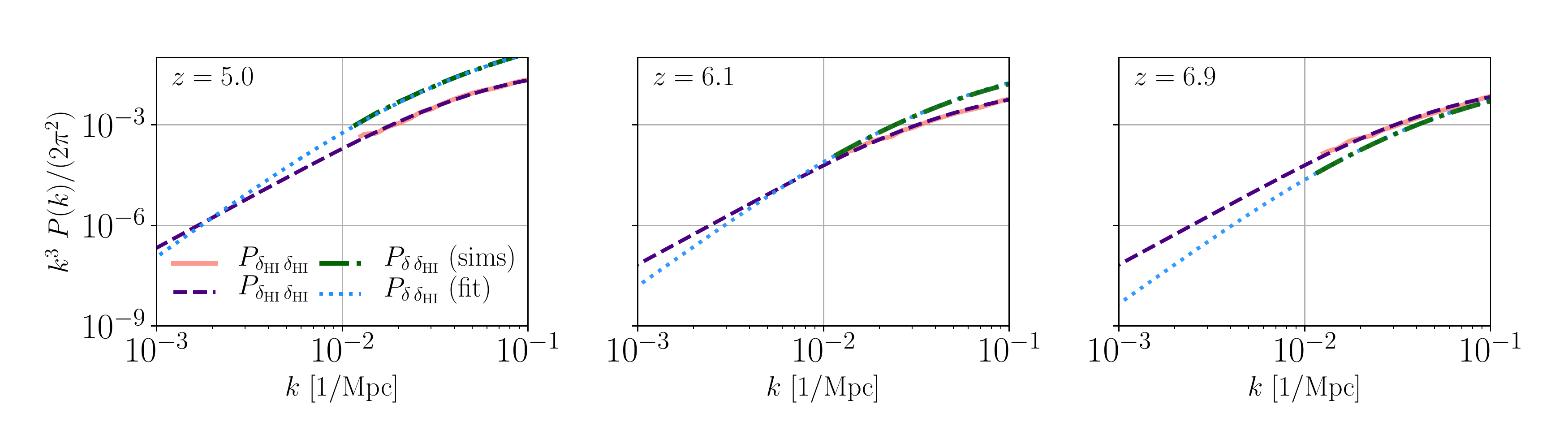}
    \vspace*{-1cm}
    \caption{The anticipated auto- and cross-correlation signals from neutral-hydrogen and density fluctuations. The light-red (purple) solid (dashed) lines correspond to auto-correlation power-spectra of the neutral hydrogen fluctuations. The green (blue) dot-dashed (dotted) lines correspond to cross-correlation of the hydrogen and the density perturbations. We extrapolate the results from the semi-numerical \texttt{21cmFAST} code to larger scales with the bubble model, introduced in Section~\ref{sec:theory}. We use the Planck 2019 cosmological parameters and the LCDM model for the matter power-spectra and fit the bubble-model parameters to the simulations. Solid and dot-dashed lines show our results from simulations while dashed and dotted lines are our results from the fits. We show the spectra at three redshift values: $z\in\{5.0,6.1,6.8\}$ which dominate the signal-to-noise in our analysis in Section~\ref{sec:forecast}.}
    \label{fig:signals-plot2}
\end{figure*}

The 21cm-temperature perturbation, in the limit detailed in Section~\ref{sec:intensity-quadrupole}, becomes $\delta_{21} \equiv n_{\rm HI} / \bar n_{\rm HI} - 1$. The bubble model then predicts\footnote{Note that Ref.~\cite{Wang:2005my} predicts spectra involving $\delta_{\rm HI} \equiv (n_{\rm HI}-\bar n_{\rm HI})/\bar n_{\rm H}$, which is normalized differently from $\delta_{21} \equiv n_{\rm HI} / \bar n_{\rm HI} - 1$ interested in here. The conversion is $\delta_{21} = \delta_{\rm HI}/(1-x_e)$. Here, Eqs.~(\ref{eqn:21-21-power}) and (\ref{eqn:21-delta-power}) have already been adjusted with appropriate powers of $(1-x_e)$.} its auto-spectrum
\begin{multline}\label{eqn:21-21-power}
    P_{\delta_{21},\delta_{21}} = [b \langle W_R(k)\rangle\ln(1-x_e) + 1]^2 P_{\delta\delta}\\ + \frac{x_e}{1-x_e}[\langle V_b\rangle \langle W_R^2(k)\rangle + \tilde P_{\delta\delta}],
\end{multline}
and its cross-spectrum with matter-density perturbation
\begin{equation}\label{eqn:21-delta-power}
    P_{\delta_{21}, \delta} = [b \langle W_R(k) \rangle \ln(1-x_e) + 1]P_{\delta\delta}.
\end{equation}
In those equations, we define the mean bubble volume
\begin{equation}
    \langle V_b \rangle \equiv \int P(R)dR\, V_b(R),
\end{equation}
where $V_b(R) \equiv 4\pi R^3/3$ is the volume of single bubble with radius $R$. We also define the volume-weighted window function as
\begin{equation}
    \langle W^n_R(k) \rangle \equiv \frac{1}{\langle V_b \rangle^n} \int P(R)dR\, V^n_b(R) W^n_R(k) \quad(n=1,2),
\end{equation}
where $W_R(k) = 3[\sin(kR) - kR\cos (kR)]/(kR)^3$ is the spherical top hat window function in the Fourier space. In addition, we define 
\begin{equation}
    \tilde P_{\delta\delta} \equiv \frac{P_{\delta\delta}\langle V_b\rangle \langle \sigma_R^2\rangle}{\sqrt{P_{\delta\delta}^2 + \langle V_b\rangle^2 \langle \sigma_R^2\rangle^2}},
\end{equation}
where $P_{\delta\delta}$ is the matter power spectrum, and
\begin{equation}
    \langle \sigma^2_R \rangle \equiv \frac{1}{\langle V_b\rangle^2} \int P(R)dR\, V^2_b(R) \sigma^2_R
\end{equation}
is the volume-weighted average of
\begin{equation}
    \sigma^2_R \equiv \int \frac{dk}{k}\, W_R^2(k) \frac{k^3}{2\pi^2} P_{\delta\delta}(k),
\end{equation}
the variance of the mean fluctuation in a spherical region of radius $R$. Eqs.~(\ref{eqn:21-21-power}) and (\ref{eqn:21-delta-power}), plus the matter power spectrum $P_{\delta\delta}$ can then be used to compute the angular power spectrum via Eq.~(\ref{eqn:angular-power-spectrum}). We model the redshift evolution of the mean ionization fraction $x_e$ with a hyperbolic tangent as
\begin{equation}\label{eq:mean_reio}
x_e(z)=\frac{1}{2}\left\{1-\tanh{\left[\frac{y(z)-y_{\rm re}}{\Delta_y}\right]}\right\}\,,
\end{equation}
where $\Delta_y$ and $y_{\rm re}$ are model parameters and $y(z)=(1+z)^{3/2}$. 
We illustrate the results of our bubble-model calculation---whose parameters are chosen to match the semi-numerical \texttt{21cmFAST} simulations---in Fig.~\ref{fig:signals-plot2}, together with the results from simulations, for a set of redshifts where the signal-to-noise of the effect is highest. In order to match the redshift evolution of the mean ionization fraction, we have chosen $y_{\rm re}=25.6$ and $\Delta_y=4.9$, which match well to the simulations at $z<7.2$. 

\section{Forecast}
\label{sec:forecast}

Here, we assess the detectability of the cross-correlation signal between the 21-cm polarization and the CMB. Upcoming 21-cm power-spectrum experiments, such as HERA~\citep{DeBoer:2016tnn} and SKA1-low~\citep{Braun:2019gdo,Bacon:2018dui}, will generate a wealth of high-precision data in the following years. These experiments are primarily designed as interferometers and the bulk of the scientific effort in modelling the involved systematics focuses on increasing the measurement quality on scales relevant to the baryon-acoustic oscillations, $0.1\,h/{\rm Mpc}<k<1.0\,h/{\rm Mpc}$. These radio interferometers are limited on large scales, however, by their minimum dish (or antenna) separation. This sets a limit on the minimum $\ell$ multipole that can be measured. The anticipated designs of HERA and SKA experimental configurations suggest that multipoles lower that $\sim10$ will not be accessible when operated as interferometers. Nevertheless, experiments like HERA can also be operated in a single-dish mode, allowing a solution to this problem~\citep[see e.g.][]{Bull:2014rha}.  

Another cause for concern is the observational systematics, which includes the beam-profile, the $1/f$ noise and the foreground wedge~\citep{Parsons:2012qh,Liu:2014bba,Liu:2014yxa} of the 21-cm experiment at hand. These have been primarily studied in the context of interferometer experiments. Encouragingly, however, phase-1 experiments like MeerKAT\footnote{\href{https://www.sarao.ac.za/gallery/meerkat}{www.sarao.ac.za}} (an SKA precursor on the planned site of SKA1-MID~\citep{Booth:2009ex}) are already taking data, albeit on low redshifts $z<3$. Nevertheless, overcoming these systematic challenges should be equally plausible for intensity mapping at higher redshifts. It is not difficult to imagine that path-finder experiments like MeerKAT will provide the road-map for using HERA (or similar precision) experiments in single-dish mode when measuring the epoch of reionization.

Finally, it is possible for the 21-cm polarization measurements to suffer unique (and major) challenges in addition to intensity. These include calibration~\citep{1996A&AS..117..137H,2013ApJ...771..105B,2019ApJ...882...58K}, foregrounds from polarized synchrotron emission~\citep{WMAP:2006rnx,2016ApJ...830...38L,2019A&A...623A..71V}, instrumental leakage of intensity into polarization~\citep{2018MNRAS.476.3051A}, Faraday rotation from various sources of magnetic fields~\citep{2010MNRAS.409.1647J,2011A&A...531A.159S,Moore:2013ip} and depolarization effects~\citep{1966MNRAS.133...67B,2020ApJ...894...38P}. While there is ongoing research concerning these systematics, their effective mitigation may be challenging for the near-future surveys. Nevertheless, since these effects are not correlated with the CMB data, the cross-correlation signal that we study in this work should be less susceptible to systematics and may prove a promising direction forward. 

\begin{figure}
    \centering
    \includegraphics[width = \linewidth]{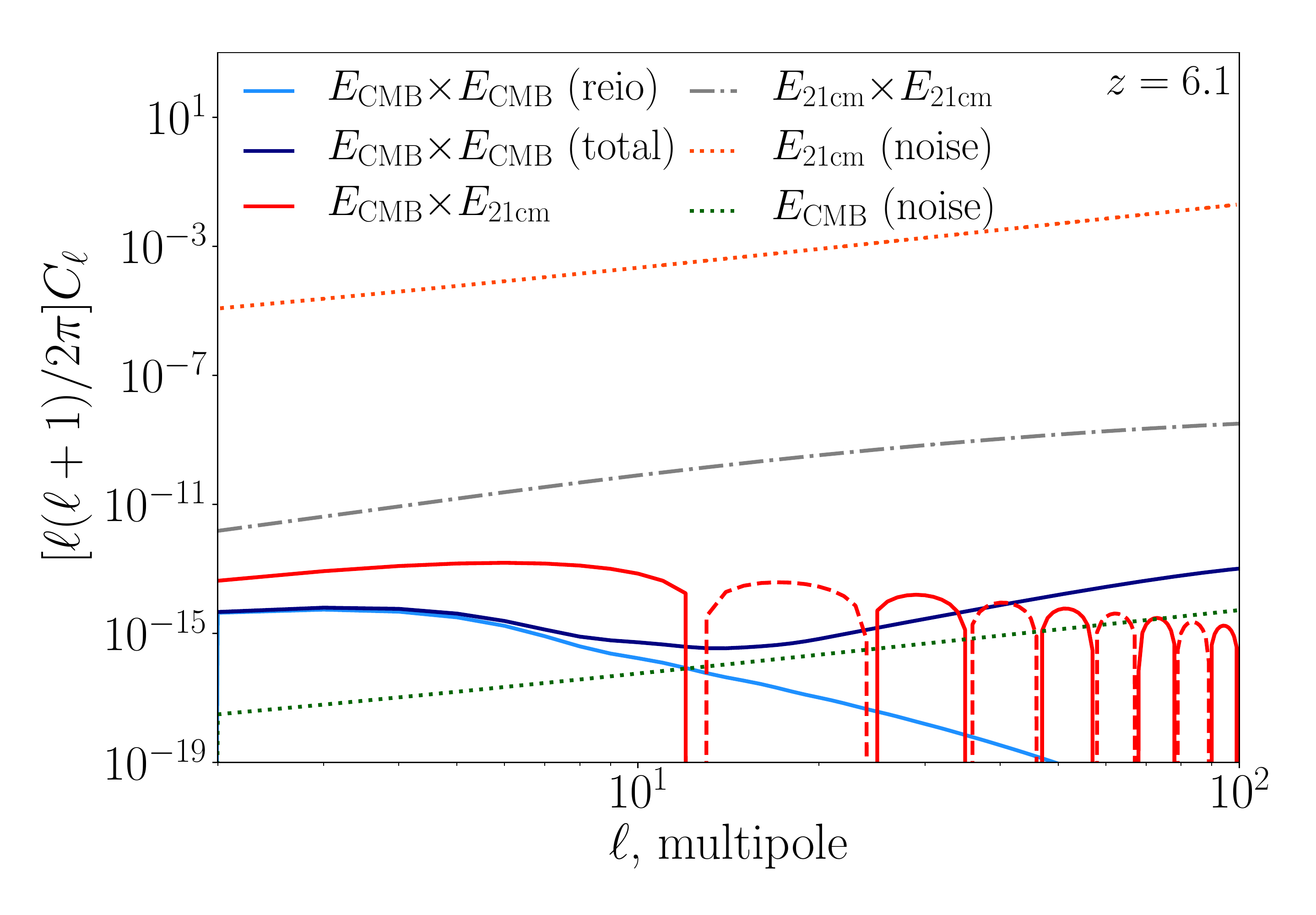}
    \vspace*{-1cm}
    \caption{CMB polarization spectrum, the anticipated polarization auto-correlation from reionization and the anticipated cross-correlation between the CMB polarization and the 21-cm. Solid dark blue line is the total CMB signal, including the reionization bump, latter shown with light blue solid line. The dot-dashed grey line is the auto-correlation of the 21-cm polarization. The red solid line is the cross-correlation signal where the negative values are shown with the red dashed line. The dotted orange line is the anticipated 21-cm polarization noise for the fiducial experiment we consider, as described in Section~\ref{sec:forecast}. The green dotted line is the Planck instrument noise. The 21-cm signal and noise were calculated at redshift $z=6.1$ and for a redshift bin of size $\Delta z=1$. We have taken $\Delta_T=1\mu$K-arcmin for demonstration.}
    \label{fig:signals-plot1}
\end{figure}

In what follows we model the anticipated noise (per multipole) from a single-dish 21-cm power-spectrum measurement as~\citep{Battye:2012tg,Pourtsidou:2015mia,Pourtsidou:2016dzn}:
\begin{equation}
    N_\ell^{E_{\rm 21}(z)}\!\!=\!\frac{\Omega_{\rm pix}T^2_{\rm sys}}{Bt_{\rm o}\bar{T}_b^2}\,{\rm exp}\left\{\ell(\ell+1)(\theta_B \sqrt{8\ln2})^2\right\}\,,
\end{equation}
where $\theta_B=\lambda/D_{\rm dish}$ is the beam full-width at half-maximum of a single dish with diameter $D_{\rm dish}$ at wavelength $\lambda=\lambda_{21}(1+z)$, $\Omega_{\rm pix}=1.13\theta_B^2$ for a Gaussian beam, $T_{\rm sys}$ is the system temperature, $B$ is the frequency bandwidth of observation, $t_{\rm o}$ is the total observation time and $\bar{T}_b(z)$ is the measured mean baryonic temperature at redshift $z$. We ignore the so-called foreground wedge~\citep[see e.g.][]{Pober:2012zz}) since, in principle, foregrounds do not lead to a loss of information (a similar approach is taken in e.g.~Ref.~\citep{Li:2018izh}), and the wedge can potentially be removed with better understanding of the  instrument~\citep{Gagnon-Hartman:2021erd,Liu:2016xzv}.  We demonstrate the anticipated noise on the 21-cm polarization in Fig.~\ref{fig:signals-plot1} along with the anticipated polarization signals from the CMB and 21-cm and the CMB intensity noise for \textit{Planck}.

We define the detection signal-to-noise (SNR) of the cross-correlation as:
\be\label{eq:SNR}
    &&{\rm SNR}^2 \\ 
    &&=\!\!\sum_{\ell\ell'}C_\ell^{E_{\rm CMB}E_{\rm 21}}\textbf{cov}^{-1}(\tilde{C}_\ell^{E_{\rm CMB}E_{\rm CMB}},\tilde{C}_{\ell'}^{E_{\rm 21}E_{\rm 21}})C_{\ell'}^{E_{\rm CMB}E_{\rm 21}}\,,\nonumber
\ee
where 
\be\label{eq:SNR2}
&&\textbf{cov}(\tilde{C}_\ell^{E_{\rm CMB}E_{\rm CMB}},\tilde{C}_{\ell'}^{E_{\rm 21}E_{\rm 21}})\\
&&\!=\!\frac{\delta_{\ell\ell'}f_{\rm sky}^{-1}}{2\ell+1}
\!\left(\tilde{C}_{\ell}^{E_{\rm CMB}E_{\rm CMB}}\tilde{C}_{\ell}^{E_{\rm 21}E_{21}}\!\!+\!\tilde{C}_{\ell}^{E_{\rm CMB}E_{21}}\tilde{C}_{\ell}^{E_{\rm CMB}E_{21}}\right).\nonumber
\ee
Here, $f_{\rm sky}$ is the sky fraction, $C_\ell^{E_{\rm CMB}E_{\rm 21}}$ is the cross-correlation between the CMB polarization and the 21-cm polarization, $C_\ell^{E_{\rm CMB}E_{\rm CMB}}$ and $C_\ell^{E_{\rm 21}E_{\rm 21}}$ are the auto-correlations of the CMB and 21-cm polarization, respectively, and spectra denoted with `tilde's are observed values, i.e. $\tilde{C}_{\ell}=C_{\ell}+N_{\ell}$. 

In what follows, we test two hypotheses. First, to access the detectability of the polarization cross-correlation signal between 21-cm  and CMB, we take a null hypothesis scenario in which there is no reionization bump signal in the CMB auto-correlation in the denominator of~Eq.~\eqref{eq:SNR} and exclude the cosmic variance from the 21-cm polarization auto-correlation. We also take the noise and signal to be diagonal in redshift bins and no signal or noise in the cross-correlation signal in the denominator of Eq.~\eqref{eq:SNR}. Second, we ask: `\textit{knowing that there is indeed a period of reionization, would we see the 21cm-CMB E mode correlation?}' For the latter, we include the reionization bump signal in the CMB auto-correlation cosmic variance. 

We consider a fiducial experiment with $T_{\rm sys}=40$~K and satisfying $\theta_B(z)=20$~arcmin at $z=6.1$. We take 5 redshift bins in the range $z\in[4,11]$ with redshift depth $\Delta z=1$, and model the CMB noise as $N_\ell^{EE}=2\Delta_T^2\exp\{\ell(\ell+1)\theta_{\rm FWHM}^2/8\ln 2\}$ where $\theta_{\rm FWHM}$ is the beam size of the CMB experiment, satisfying $\theta_{\rm FWHM}\simeq5$~arcmin for \textit{Planck}. For our first hypothesis, we find such an experiment can reach detection ${\rm SNR}=2.1$ with a dedicated 2 years of observation over the full sky, if cross-correlated with a CMB experiment with thermal noise $\Delta_T=1\mu $K-arcmin. Note that this thermal noise for the CMB experiment is lower than \textit{Planck}. While upcoming CMB experiments such as SO and CMB-S4 will lower the thermal noise on small scales significantly, improving the fidelity of large-scale measurements ($\ell<30$) requires satellite experiments that do not suffer the atmospheric noise. A promising next-generation experiment was proposed in~Ref.~\citep{Basu:2019rzm}, which may play a crucial role in measuring the cross-correlation signal we study in this paper. For the null hypothesis the SNR depends on survey specifications as 
\begin{equation}\label{eq:SNR_main}
{\rm SNR}\simeq2.1 {f_{\rm sky}}\!\left(\frac{20'}{\!\theta_B^*}\!\right)^2\!\!\left(\frac{1\mu{\rm K}'}{\Delta_T}\right)^2\!\!\left(\frac{40{\rm K}}{T_{\rm sys}}\right)^2\frac{t_0}{2{\rm yrs}}\,,
\end{equation}
where $\theta_B^*$ is the beam at $z_*=6.1$. Here, we have taken the minimum multipole as $\ell_{\rm min}=2$. For our second hypothesis, we find the coefficient of Eq.~\eqref{eq:SNR_main} degrade by over an order of magnitude to ${\rm SNR}\simeq0.017$ due to the large cosmic-variance of the reionization bumb. For the first hypothesis, we find that the SNR depends on the choice of $\ell_{\rm min}$ as ${\rm SNR}\propto\exp\{-4\ell_{\rm min}/5\}$, suggesting that the measurement of the largest scales will be crucial for detection. We find the decrease with increasing multipoles is less rapid for the second hypothesis, where the SNR depends on choice of $\ell_{\rm min}$ as ${\rm SNR}\propto\exp\{-0.3\ell_{\rm min}\}$. Since reaching any of these specifications remains challenging for the near-future experiments, we conclude detecting the cross-correlation signal will be difficult, but not impossible with dedicated next-generation surveys.

\section{Conclusion}
\label{sec:conclusion}

In this paper, we propose to enhance the detectability of the 21-cm polarization by cross-correlation with the CMB polarization. We present the theoretical procedure to compute this cross-correlation from reionization physics, which is then modeled by the bubble model with parameters calibrated to the simulation results of \texttt{21cmFAST}. We demonstrate the prospect of detection by performing a basic noise analysis, and give the SNR as a function of observational parameters.

In the noise analysis, we choose to test two closely related, but subtly different, null hypotheses. In the first scenario, we assume that there is no reionization, and subsequently no 21-cm or reionization-era CMB polarization. In the second scenario, we assume that there is no 21-cm polarization, but the reionization-era CMB polarization is present. In each scenario, we forecast how strongly will the null hypotheses be disfavored by the observation of 21cm-CMB polarization correlation.

We find that, with very generous assumptions on the 21-cm observational systematics, the synergy of ambitious next-generation 21-cm and CMB missions could make a detection in the first scenario, while the second scenario will still remain out of reach. In the first scenario, the observation at large angular scales is crucial, as that is where the correlation gains most of its contribution.

Future work could improve on the reionization modelling with larger numerical simulations; include the effect of redshift-space distortion; discuss the 21cm-radiation field before the saturation of heating; provide a more realistic account for the 21-cm observational systematics; devise strategies to mitigate foregrounds unique to the 21-cm polarization measurement, especially the Faraday rotation induced by the magnetic fields.

\begin{acknowledgments}
This work was supported by NSF Grant No.\ 1818899 and the Simons Foundation. This work was performed in part at Aspen Center for Physics, which is supported by National Science Foundation grant PHY-1607611, and was partially supported by a grant from the Simons Foundation.  SCH is supported by the Horizon Fellowship from Johns Hopkins University

\end{acknowledgments}

\appendix

\section{Derivation of the quadrupole tensor and its TAM coefficients}
\label{app:quadrupole-fourier}

When the observed anisotropy at $(\vec x, \eta)$ is determined by the value of some field $\phi$ on the emission shell $\vec x_e \equiv \vec x + (\eta - \eta_e)\hat u$ at conformal time $\eta_e$,
\begin{equation}
    \Theta(\hat u, \vec x, \eta) = \phi[\vec x + (\eta - \eta_e)\hat u, \eta_e],
\end{equation}
the associated quadrupole tensor is
\begin{equation}
    t_{ab}(\vec x, \eta) \equiv \int d^2u\, (3u_a u_b-\delta_{ab})\phi[\vec x + (\eta - \eta_e)\hat u, \eta_e].
\end{equation}
In Fourier space, this relation takes the form
\begin{align}
    \tilde t_{ab}(\vec k, \eta) &= \int d^2u\, (3u_a u_b-\delta_{ab})\int d^3 x\, \exp(-i\vec k\cdot \vec x)  \nonumber \\ &\quad \quad \quad \quad \times \phi[\vec x + (\eta - \eta_e)\hat u, \eta_e] \nonumber \\
    & = \tilde \phi(\vec k, \eta_e) \int d^2u\, (3u_a u_b-\delta_{ab}) \exp[i\vec k\cdot \hat u(\eta-\eta_e)] \nonumber \\
    & = -12\pi j_2[k(\eta-\eta_e)]\left(\hat k_a \hat k_b -\frac{\delta_{ab}}{3}\right)\tilde \phi(\vec k, \eta_e).
\end{align}
Here, in the second equal sign, we have used the shift formula of the Fourier transform. In the third equal sign, the integral can be evaluate by taking $\vec k = k \hat z$, and then be restored to its general form via symmetry.

\section{Transfer functions}
\label{app:tam-transfer}

Eqs.~(8) and (104) of Ref.~\cite{Dai:2012bc} can convert the Fourier amplitudes to the TAM coefficients, giving
\begin{equation}
    t_{kJM}(\eta) = 4\pi\sqrt6 j_2[k(\eta-\eta_e)]\phi_{kJM}(\eta_e).
\end{equation}

For the 21-cm signal and the CMB, let $\phi$ be $\delta_{21}$ and $-\Phi/3$, respectively. We have
\begin{align}
    t^{21,\nu_o}_{kJM}(\eta) &= \mathcal T_{21,\nu_o}(k; \eta, \eta_e) \delta^{21}_{kJM}(\eta_e);\\
    t^{\rm CMB}_{kJM}(\eta) &= \mathcal T_{\rm CMB}(k; \eta, \eta_e) \delta_{kJM}(\eta_e),
\end{align}
where
\begin{align}
    \mathcal T_{21,\nu_o}(k; \eta, \eta_e) &= 4\pi \sqrt6 j_2[k(\eta-\eta_e)];\\ 
     \mathcal T_{\rm CMB}(k; \eta, \eta_e) &= -\frac{2\pi\sqrt6 \Omega_m H_0^2}{k^2 a(\eta_e)} j_2[k(\eta-\eta_*)].
\end{align}
In the second formula we used the relation $\tilde\Phi(\vec k,\eta)=3\Omega_m H_0^2\tilde \delta(\vec k, \eta)/[2k^2a(\eta)]$ and the fact that $\tilde\Phi(\vec k,\eta)$ is approximately constant deep in the matter-dominated era.

\end{document}